\documentclass[11pt,a4paper,twocolumn]{article}
\usepackage{graphicx}

\usepackage{bm}

\begin{document}

\title{Superheavy Elements: \\
Beyond the 7th Period in the Periodic Table}
\author{KOUICHI HAGINO \\
DEPARTMENT OF PHYSICS, TOHOKU UNIVERSITY, JAPAN
}
\date{}

\maketitle

{\bf ABSTRACT}  \\

What is the heaviest element? 
In order to address this question, 
the elements up to $Z=118$ have been synthesized by now by using heavy-ion 
fusion reactions. 
This has completed the 7th period in the periodic table of elements, 
and new attempts have been commenced in aiming at syntheses of 
the elements in the 8th period. 
In this article, we review the current status and future challenges 
in the research 
field of superheavy elements, putting some 
emphasis on perspectives from 
the nuclear reaction theory. 
\\

{\bf INTRODUCTION} \\

The elements heavier than Plutonium (the atomic number 
$Z=94$) are all unstable and do not 
exist on the Earth. Yet, one can artificially synthesize them using nuclear 
reactions. There have been continuous efforts since the 1950s (see e.g., 
Fig. 1 in Ref. \cite{ST12}), and the elements up to $Z=118$ have been 
synthesized by now. Out of these 118 elements, four new elements, 
$Z=113$ (Nihonium, Nh), $Z=115$ (Moscovium, Mc), $Z=117$ (Tennessine, Ts), 
and $Z=118$ (Oganesson, Og) are the most recent ones, 
which were added to the 
periodic table of elements in 2016 \cite{PAC16}, 
completing the 7th period in the 
periodic table (see Ref. \cite{Zhou17} for an interesting 
article on the Chinese characters for these elements). 
It is worth mentioning that Nihonium is the first element which was named 
after an Asian country. 

The transactinide elements, that is, the elements with $Z\geq 104$, are 
referred to as superheavy elements. 
Those superheavy elements are interesting many-body systems, 
as they can also be viewed as 
quantum laboratories under 
the influence of a strong Coulomb field generated by 
many protons in their atomic nuclei. 
In fact, for the following reasons, they offer an ideal opportunity 
to address a fundamental question of many-body physics:  
how does a quantum many-body system 
behave under the influence of a strong Coulomb field? 
Firstly, in atomic 
nuclei in the superheavy elements, i.e., in superheavy nuclei, 
there is a strong interplay between the strong and the electromagnetic 
interactions, making superheavy nuclei unique many-body systems. 
That is, while the strong interaction plays a dominant role in usual nuclei, 
both the strong and the Coulomb interactions contribute in a similar way  
in superheavy nuclei. Because of this, quantum effects appear more prominent 
in superheavy nuclei than in usual nuclei. 
The effect of shell correction on a fission barrier is a 
typical example \cite{Moller15}. 
Furthermore, the electric dipole moment (EDM), which is intimately related 
to the fundamental symmetries such as CP symmetry, is enhanced in heavy 
elements (for alkali atoms, for instance, the enhancement factor is 
scaled as $Z^3$ \cite{Gingers04}), providing a good motivation to study 
heavy and superheavy elements. 
Secondly, it is known in chemistry that the periodic table can be  
perturbed due to the strong Coulomb interaction among electrons, as can 
be seen 
in lantanides and actinides. In superheavy elements, 
the relativistic effect becomes prominent, and 
the periodic table may further be disturbed significantly. 

Heavy-ion fusion reactions at energies around the 
Coulomb barrier have been used as 
a standard tool 
to synthesize those superheavy elements \cite{Hofmann00,Hamilton13}. 
Apparently, it is indispensable 
to understand its reaction dynamics 
in order to synthesize efficiently new elements, going beyond the 
7th period in the periodic table (see Ref. \cite{Hofmann18} 
for new criteria for a discovery of a new element). However, 
the fusion reaction in the superheavy region is nothing more than 
the dynamics of many-body systems under the strong Coulomb field, and 
there still remain many challenges. 
In this article, we shall 
review the current status and future perspectives of the nuclear reaction studies 
for superheavy elements.
We refer to  
two recent articles, Refs. \cite{Witek18,Giuliani18}, 
for complementary reviews of superheavy elements from 
the perspectives of nuclear structure theory. 
\\

{\bf HEAVY-ION FUSION REACTIONS FOR SUPERHEAVY ELEMENTS} \\

\begin{figure}[bt]
\begin{center}
\includegraphics[width=0.9\linewidth]{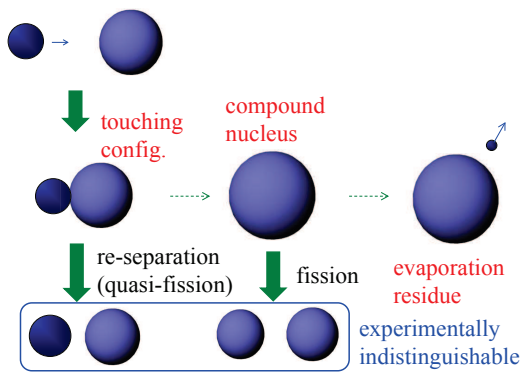}
\end{center}
\caption{A schematic illustration of a heavy-ion fusion reaction to 
form a superheavy element. The re-separation process without forming 
a compound nucleus (quasi-fission) 
and a fission decay of a compound nucleus cannot be 
separated experimentally, and the formation of a compound nucleus 
is identified by detecting evaporation residues.}
\end{figure}

{\bf Overview of the reaction process}\\
Nuclear fusion is a reaction in which two nuclei combine together to form 
a larger nucleus, which is referred to as a compound nucleus. 
Figure 1 illustrates schematically a nuclear fusion to form a superheavy 
element. In the first phase of reaction, two nuclei approach to the touching 
configuration. There is a potential barrier between the two nuclei, due to a 
cancellation between the long range Coulomb interaction and a short range 
attractive nuclear interaction, which has to be overcome in order to reach the 
touching configuration. For medium-heavy systems, a compound nucleus is 
formed almost automatically once the touching configuration 
is achieved \cite{HT12}. In contrast, in the superheavy region, there is a 
huge probability for the touching configuration to reseparate due to a 
strong Coulomb repulsion between the two nuclei. 
Furthermore, even when a compound nucleus is formed with a small probability, 
it decays most likely by fission, again due to the strong 
Coulomb interaction. A complication is that quasi-fission 
characteristics significantly overlap with fission of the compound nucleus, 
and a detection of fission events itself does not guarantee a formation of the 
compound nucleus. Therefore, one really 
needs to detect evaporation residues, that is, 
those extremely rare events in which a compound nucleus is 
survived against fission. 

\begin{figure}[bt]
\begin{center}
\includegraphics[width=0.75\linewidth]{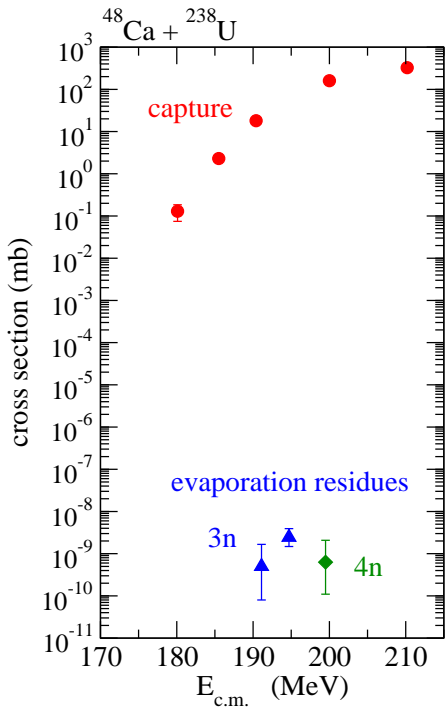}
\end{center}
\caption{The measured evaporation residue cross sections 
as a function of the bombarding energy in the center of mass frame 
for the 
$^{48}$Ca+$^{238}$U reaction leading to the formation of Cn ($Z=112$) element. 
The filled circles show the capture cross sections \cite{Kozulin14} 
to form the touching 
configuration shown in Fig. 1. The filled triangles and diamonds show 
the evaporation residue cross sections \cite{Oganessian04}, 
for which the former and the latter 
correspond to the 3$n$ (emission of 3 neutrons) and the 
4$n$ (emission of 4 neutrons) channels, respectively. }
\end{figure}

As an example, Fig. 2 shows the measured cross sections for 
the $^{48}$Ca+$^{238}$U reaction 
forming the Cn ($Z=112$) element. The filled circles show the capture 
cross sections \cite{Kozulin14} 
to form the touching configuration shown in Fig. 1. 
On the other hand, the filled triangles and diamonds show the evaporation 
residue cross sections \cite{Oganessian04}, for which the former and the 
latter correspond to the process of emission of 3 and 4 neutrons,  
respectively. One can see that the evaporation residue cross sections 
are smaller than the capture cross sections by about 11 orders of magnitude. \\

{\bf Theoretical modelings} \\
Based on the time-scale of 
each process, the formation process of evaporation residues  
can be conceptually divided into a sequence of the following 
three processes (see Fig. 1). 
The first phase is a process in which two separate nuclei form the touching 
configuration after overcoming the Coulomb barrier. 
Here, the couplings of the relative motion to several nuclear 
collective excitations in the colliding nuclei 
as well as several transfer processes play an important role 
\cite{HT12}. 
After two nuclei touch with each other, a huge number of nuclear intrinsic 
motions are activated 
and the energy for the relative motion of the colliding nuclei 
is quickly dissipated to internal energies. 
Because of the strong Coulomb interaction, 
the touching configuration appears outside the saddle configuration of a fission 
barrier, and 
thus a compound nucleus is formed only after the fission barrier 
is thermally activated whereas most of events go to 
quasi-fission. 
In order to describe this process, 
a Langevin dynamics has been developed \cite{Abe02,fbd03,Aritomo04,ZG15}, 
although the dinuclear system model \cite{Adamian98} has also been used. 
The third process is a statistical decay of the compound 
nucleus \cite{KEWPIE2}, with strong competitions  
between fission and particle emissions (i.e., evaporations). 
Here, the fission barrier height is one of the most important parameters 
which significantly affect evaporation residue cross sections \cite{Boilley16}. 

For a given partial wave $\ell$, the probability for a formation of 
an evaporation residue is given as a product of the probability for each 
of the three processes, that is, 
\begin{equation}
P_{\rm ER}(E,\ell)=T_\ell(E)P_{\rm CN}(E,\ell)W_{\rm sur}(E^*,\ell),
\end{equation}
where $E$ and $E^*$ are the bombarding energy in the center of mass 
frame and the excitation energy of the compound nucleus, respectively. 
As has been mentioned, there is no way to access experimentally 
the formation of the compound nucleus and no experimental data 
are available for $P_{\rm CN}$. 
This causes large ambiguities in 
theoretical calculations. 
An important theoretical challenge is then to reduce theoretical uncertainties 
in model calculations, especially for $P_{\rm CN}$, 
and make reliable predictions for evaporation residue 
cross sections. \\

\begin{figure}[bt]
\begin{center}
\includegraphics[width=0.9\linewidth]{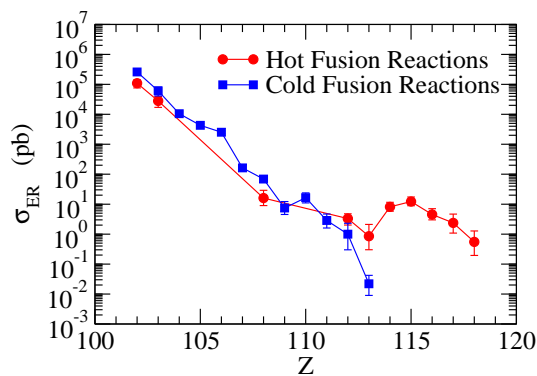}
\end{center}
\caption{The measured evaporation residue cross sections 
as a function of the atomic number $Z$ of a compound nucleus. 
The filled circles denote the results of the hot fusion reactions, 
in which $^{48}$Ca 
nucleus is used as a projectile. The maximum of a sum of the 3$n$ and 4$n$ 
cross sections are shown for each $Z$. 
The filled squares show the 
results of the cold fusion reactions, in which the $^{208}$Pb or $^{209}$Bi 
nuclei 
are used as a target.  Here, the maximum of the 1$n$ cross sections 
are shown for each $Z$. 
The experimental data are taken from Refs. \cite{NRV,Oganessian13,Morita12}.}
\end{figure}

{\bf Hot fusion and cold fusion reactions} \\
Since a formation of evaporation residues is a very rare process, 
it is important to choose appropriate combinations of the projectile and 
the target nuclei in order to efficiently synthesize superheavy elements. 
For this purpose, two different experimental strategies have been employed.  
One is the so 
called ``cold fusion'' reactions, for which the compound nucleus is formed 
with relatively low excitation energies so that the competition 
between neutron emissions and fission can be 
minimized, thus maximizing $W_{\rm sur}$ in Eq. (1) \cite{Hofmann00,Hamilton13}. 
To this end, $^{208}$Pb and $^{209}$Bi are used for the target nuclei. 
An advantage of this strategy is that 
alpha decays of the evaporation residues end up in the known region of 
nuclear chart, and thus superheavy elements can be identified unambiguously. 
Nihonium was synthesized at RIKEN using this strategy \cite{Morita12}. 
The other strategy is the so called ``hot fusion'' reactions, 
for which more asymmetric combinations of the projectile and 
the target nuclei are used as compared to the cold fusion reactions, 
so that the formation probability of the compound nucleus, 
$P_{\rm CN}$ in Eq. (1), can be optimized. 
For this purpose, the neutron-rich double magic nucleus $^{48}$Ca has 
been used as a projectile \cite{Hamilton13,Oganessian15}. 
This strategy has been successfully employed by the experimental group at 
Dubna, led by Oganessian, to synthesize superheavy 
elements up to $Z=118$ (see e.g., Fig. 2). 
See also Ref. \cite{Gates18} for the very first direct measurement of the mass 
numbers of superheavy elements synthesized by the hot fusion reactions. 

Figure 3 compares the measured evaporation residue cross sections due to 
the hot fusion reactions (the filled circles) with those due to 
the cold fusion reactions (the filled squares). 
For the cold fusion reactions, the cross sections drop rapidly as a 
function of $Z$ of the compound nucleus. It would therefore be difficult 
to go beyond Nihonium using this strategy. In contrast, for the hot fusion 
reactions, the cross sections remain relatively large between $Z=113$ and 118. 
This is due to the fact that the compound nuclei formed are in the 
proximity of the predicted island of 
stability \cite{island1,island2} and/or an increase of dissipation at 
high temperatures \cite{Loveland14}, both of which increase the survival 
probability, $W_{\rm sur}$. \\

{\bf ROLE OF DEFORMATION IN HOT FUSION REACTIONS} \\

\begin{figure}[bt]
\begin{center}
\includegraphics[clip,width=0.9\linewidth]{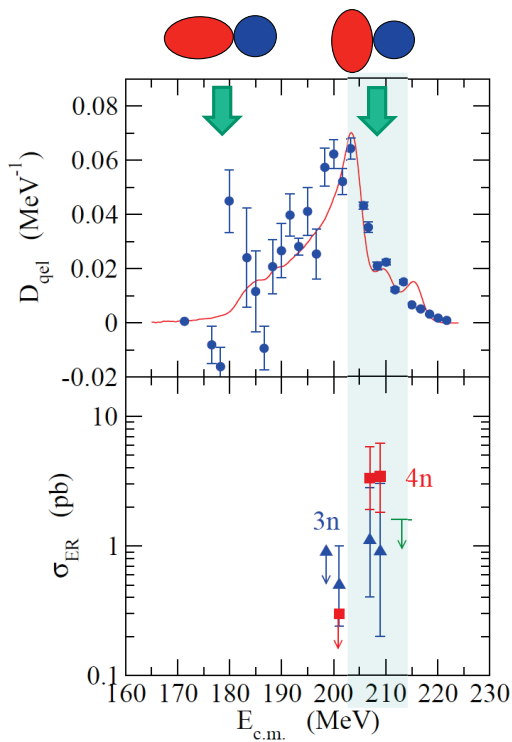}
\end{center}
\caption{(Upper panel) The experimental quasi-elastic 
barrier distribution, $D_{\rm qel}$, 
for the $^{48}$Ca+$^{248}$Cm system \cite{Tanaka18}. 
The solid line shows the result of a coupled-channels calculation 
which includes the deformation of $^{248}$Cm as well as a transfer coupling. 
The shaded region corresponds to the so called the side collision. 
(Lower panel) The experimental evaporation residue cross 
sections for this system taken from Refs. \cite{Oganessian04,Hofmann12}.}
\end{figure}

\begin{figure}[bt]
\begin{center}
\includegraphics[clip,width=0.9\linewidth]{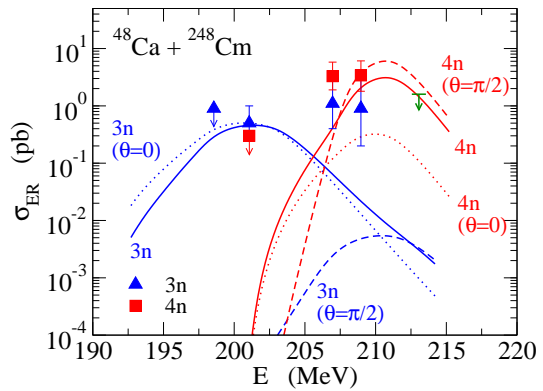}
\end{center}
\caption{
The evaporation residue cross sections for the 
$^{48}$Ca+$^{248}$Cm system obtained with the extended 
fusion-by-diffusion model \cite{Hagino18}.}
\end{figure}

{\bf Quasi-elastic barrier distribution} \\
In the hot fusion reactions, by fixing the projectile nucleus to be 
$^{48}$Ca, the target nuclei are found in the actinide region, in which 
the nuclei are well deformed in the ground state. An interesting 
and important question to ask 
is how deformation of the target nuclei affects evaporation 
residue cross sections. 
When a target nucleus is deformed, 
a single Coulomb barrier in the entrance channel 
is replaced by a distribution of a multitude of 
Coulomb barriers, since 
the height of the Coulomb barrier 
depends on the orientation angle of the target nucleus. 
A way how the barrier heights are distributed can be studied 
by measuring the so called
quasi-elastic barrier distribution, $D_{\rm qel}$, which is defined as the first 
energy derivative of the ratio of the quasi-elastic cross sections 
to the Rutherford cross sections at backward angles \cite{Timmers95,HR04}. 
Very recently, such measurement 
was carried out for the $^{48}$Ca+$^{248}$Cm system 
by Tanaka et al. \cite{Tanaka18}. 
Fig. 4 shows the experimental data together with the result of a 
coupled-channels calculation, which 
takes into account the deformation of the target nucleus, $^{248}$Cm. 
The figure clearly indicates that 
the maximum of the evaporation residue cross sections is obtained with 
the so called side collision, that is, the configuration in which the 
projectile approaches from the shorter axis of the target nucleus 
(with the orientation angle of $\theta=\pi/2$ with respect to the 
beam direction). 
We mention that 
this is a nice confirmation of the notion of compactness proposed by 
Hinde et al. \cite{Hinde95}, who argued that the side collision leads to 
a compact touching configuration for which the effective barrier height 
for the diffusion process is low, enhancing the formation 
probability, $P_{\rm CN}$. 
This notion has further been confirmed theoretically \cite{Hagino18} 
using an extended version of the fusion-by-diffusion model, which takes into 
account the deformation effect of the target nucleus on the injection point for the diffusion 
process (See Fig. 5). 

Another important aspect of the measurement of Tanaka et al. is 
that it provides good information on capture cross sections. 
Capture cross sections can in principle 
be measured experimentally, but often the presence of 
deep-inelastic collisions complicates their experimental determination. 
The quasi-elastic cross sections measured by Tanaka et al. are almost 
free from the deep-inelastic component \cite{Tanaka18}, and thus 
the capture cross sections, or the capture probability $T_\ell$ in Eq. (1), 
constructed from the measured quasi-elastic 
cross sections, are cleaner than those from direct measurements. 
This will be helpful in reducing theoretical uncertainties in 
modeling the formation process of evaporation residues. 

{\bf A remaining theoretical challenge} \\
Even though the quasi-elastic barrier distribution has nicely 
demonstrated the role of deformation in synthesizing superheavy elements, 
there still remains a theoretical challenge concerning hot fusion reactions 
with a deformed target. That is, it has yet to clarify how the shape of 
the di-nucleus system evolves towards a compound nucleus. As we have mentioned, 
the di-nucleus system is rapidly heated up after the touching, which 
will reduce 
several quantal effects such as nuclear deformation. 
However, nuclear deformation would persist for a while given that the notion 
of compactness is correct as has been indicated by the quasi-elastic 
barrier distribution. 
In order to clarify this, one would need to develop a microscopic 
dynamical theory, in which the heat-up process can be described in 
a consistent manner from the approaching phase 
to the formation of a compound nucleus. 
The shape of the whole system would then be determined 
self-consistently at each time of evolution. 
A candidate for such theory is the one 
developed by Mukamel et al. in the 80s 
in the context of deep-inelastic collisions \cite{Mukamel81}, even though 
any practical calculation has not been carried out based on this theory. \\

{\bf TOWARDS $Z=119$ AND 120} \\

The heaviest element synthesized so far is $Z=118$ (Oganesson). 
To go beyond this and synthesize the elements $Z=119$ and $Z=120$ with hot fusion reactions with the $^{48}$Ca projectile, 
one would need Es ($Z=99$) and Fm ($Z=100$) targets.  However, these elements are both 
short lived and are not available with sufficient amounts to perform fusion 
measurements \cite{Dullmann17}. Heavier projectile nuclei, such as 
$^{50}$Ti, $^{51}$V, and $^{54}$Cr, would then have to be used instead of $^{48}$Ca. 
For instance, a new measurement campaign 
has already been started at RIKEN to synthesize the element 119  
using the $^{51}$V + $^{248}$Cm reaction \cite{RIKEN}.  

An important issue here is to asses how much evaporation residue cross sections are affected if 
a projectile nucleus other than $^{48}$Ca is used. 
One can consider the following two effects. Firstly, while $^{48}$Ca is a 
double magic nucleus, 
$^{50}$Ti, $^{51}$V, and $^{54}$Cr are open shell nuclei with valence 
nucleons outside the $^{48}$Ca core. In the approaching phase, $^{48}$Ca could 
come closer to a target nucleus with less friction 
as compared to the other heavier projectile nuclei \cite{Satou06}. 
At the same time, the resultant compound nuclei would be at a larger excitation energy 
with the $^{48}$Ca projectile 
as compared to the heavier projectiles. The former reduces 
evaporation residue cross sections while 
the latter enhances the cross sections when the heavier projectiles are used. 
Secondly, reactions with the heavier projectiles are 
less asymmetric than those with $^{48}$Ca. 
This leads to a higher effective barrier for the diffusion process (that is, 
the second phase in the reaction process shown 
in Fig. 1), reducing evaporation residue cross sections. 

The net effect will be a combination of these effects. Among them, the effect of friction in the approaching phase could be best studied with a microscopic theory such as the Time-Dependent 
Hartree-Fock (TDHF) method. By combining results of a TDHF calculation 
and the Langevin 
dynamics, one would be able to discuss how much evaporation residue cross 
sections are reduced (or enhanced) when the heavier projectiles 
are used instead of the 
$^{48}$Ca nucleus \cite{Sekizawa19}.  
\\

{\bf TOWARDS THE ISLAND OF STABILITY} \\

One of the main motivations to study superheavy elements, in addition to 
synthesizing new elements, is to look for the island of stability, which was 
theoretically predicted some 50 years ago \cite{island1,island2}. 
Heavy nuclei in the transactinide region are unstable against 
alpha decay and spontaneous fission, but the shell effect due to magic 
numbers can stabilize a certain number of nuclei in that region. 
The predicted proton and neutron magic numbers are $Z=114$ and 
$N=184$ \cite{island1,island2}, 
respectively, and the region around these magic numbers is 
referred to as the island of stability. 
A more modern Hartree-Fock calculation has also predicted 
$(Z,N)=(114,184), (120,172)$, and (126,184) 
for candidates for the next double magic nucleus 
beyond $^{208}$Pb \cite{Bender99}. 

The heaviest Fl element ($Z=114$) synthesized so far 
is $^{289}_{175}$Fl, which was synthesized using the $^{48}$Ca+$^{244}$Pu 
hot fusion reaction \cite{Oganessian04-2}. 
Notice that 9 more neutrons are needed in order to reach the predicted 
magic number, $N=184$. 
This implies that neutron-rich beams are indispensable in order to 
reach the island of stability. 
An experimental challenge towards this direction 
is how to deal with low intensity 
of such beams. On the other hand, from a theoretical point of view, 
the reaction mechanism of fusion of neutron-rich nuclei is quite complex and 
has not yet been clarified completely \cite{Canto06}. 
In particular, a simultaneous treatment of fusion, breakup, and transfer 
processes has yet to be developed \cite{Choi18}. 
Another possibility, besides fusion, to reach the island of stability 
is to use a multi-neutron transfer reaction with neutron-rich 
beams \cite{Zagrebaev08}. 
Apparently more studies are needed, both experimentally and 
theoretically, in order to find optimum reactions and experimental 
conditions, including reaction systems, to reach most efficiently 
the island of stability. 
Of course, studies of structure of neutron-rich nuclei 
also make an important ingredient for this purpose. 
\\

{\bf ASTROPHYSICAL PERSPECTIVES} \\

It is worth mentioning that an investigation of nuclear reactions 
of neutron-rich nuclei discussed 
in the previous section may also help in clarifying an important 
question of modern science: how and where were heavy 
elements created in the universe?. 
This is concerning the r-process nucleosynthesis, 
whose pathway is through 
the neutron-rich region in nuclear chart. 
A recent simultaneous detection of gravitational and 
electromagnetic waves from a neutron star merger event GW170817 
has confirmed that 
mergers of neutron stars are important sites of 
r-process nucleosynthesis \cite{Tanaka2017,Hotokezaka2018}. 
However, there still have been many unknown features in the r-process 
nucleosynthesis. One of the key issues is the role of fission of neutron-rich nuclei. 
When 
heavy nuclei are created during the nucleosynthesis, those nuclei 
decay by fission, producing lighter fragments which may be 
recycled for nucleosynthesis \cite{Goriely13,Kajino17,Giuliani18-2}. 
That is, fission will determine the end point of 
r-process nucleosynthesis. 
However, fission is a typical example of 
large amplitude collective motions, in which 
the shape of a quantum many-body system changes largely, 
and its microscopic understanding has not yet been reached. 
In order to clarify the role of fission in the r-process 
nucleosynthesis, 
a study of nuclear reactions for superheavy elements could play an 
important role, since superheavy elements detected in such reactions 
are evaporation residues, which are survived against fission. 
That is, by combining nuclear physics and astrophysics 
together, with fission as an important key word, one would be able to achieve 
a comprehensive understanding of the synthesis of heavy and superheavy 
elements both in the universe and in laboratories. 
This will certainly be an important future direction in the research 
field of superheavy elements. 
\\

{\bf OUTLOOK} \\

How does a strong Coulomb field affect behavior of quantum many-body 
systems? How and where were heavy elements around us created in the 
universe? These are important questions in the research field of superheavy 
elements. Those questions involve 
many research fields, 
not only nuclear physics and astrophysics 
but also chemistry. Nucleonic many-body problems in nuclear physics 
and electronic many-body problems in chemistry share similar problems 
to each other. Moreover, understanding electronic structure of heavy and 
superheavy elements plays an important role in understanding opacity 
of electromagnetic radiations from r-process nucleosynthesis. 

The 7th period in the periodic table of elements has now been completed. 
Going beyond the 7th period to proceed to the 8th period is a real challenge 
now. Given this situation, 
it would be extremely useful to 
develop a multidiciprinary science 
for superheavy elements, including physics, astronomy, and chemistry. 
By combining those research fields together, 
it is likely that we will be able to answer the fundamental questions 
of superhevy elements and 
achieve their comprehensive understanding from wide perspectives. 

{\bf Acknowledgments:}
The author would like to thank Y. Abe, Y. Aritomo, and T. Tanaka 
for useful discussions.

\end{document}